\documentclass[preprint,prb]{revtex4}
\usepackage{amsmath}
\usepackage{amssymb}
\usepackage{amsthm}

\def\v#1{{\bf#1}}
\newcommand{\bff}{\boldsymbol{f}}

\newcommand{\bfh}{\boldsymbol{h}}

\newcommand{\bfchi}{\boldsymbol{\chi}}

\begin{document}
\title{Instantaneous fields in classical electrodynamics}
\author{Jos\'e A. Heras}
\email{heras@phys.lsu.edu}
\affiliation{Departamento de F\'\i sica, E.\ S.\ F.\ M., Instituto Polit\'ecnico Nacional, M\'exico D.\ F.\ M\'exico and Department of Physics and Astronomy, Louisiana State University, Baton Rouge, Louisiana 70803-4001, USA}

\begin{abstract} 
In this paper we express the retarded fields of Maxwell's theory in terms of the instantaneous fields of  a Galilei-invariant electromagnetic and we find the  vector function $\bfchi_{L}$ whose spatial and temporal derivatives transform the instantaneous fields into the retarded ones. We conclude that the instantaneous fields can formally be introduced as unphysical objects into classical 
electrodynamics which can be used to find the physical retarded fields
\end{abstract}
\pacs{03.50.De, 03.50.Kk, 41.20.Cv}
\maketitle

The possible coexistence of both instantaneous and retarded electromagnetic interactions 
has recently been discussed [1-5]. The idea of this coexistence is partially motivated by the striking result that the  scalar potential in the Coulomb gauge is an instantaneous quantity. This result was discussed some years ago by Brill and Goodman [6] and by Gardiner and Drummond [7] and more 
recently by Rohrlich [8], Jackson [9] and the author [10]. The conclusion has been reiterative: the retarded fields derived from the 
Lorenz-gauge potentials are the same retarded fields obtained from the Coulomb-gauge potentials, despite the instantaneous character of the Coulomb-gauge scalar potential. The electromagnetic theory is {\it complete} as usually expressed [5] and there is no necessity of introducing instantaneous fields into this classical theory. The physical fields are always retarded, except in the strict static limit. 

However, there is  a {\it formal} connection between instantaneous and retarded fields which is certainly 
unavoidable. The argument is as follows. Given a confined source $\bff(\v x,t)$ we can always construct simultaneously  (i) a retarded field $\v F(\v x,t)$ satisfying the wave equation: $\Box^2\v F(\v x,t)={\bff}(\v x,t)$ where  $\Box^2\equiv\nabla^2-(1/c^2)\partial^2/\partial t^2$ 
and (ii) an instantaneous field $\widetilde{\v F}(\v x,t)$ satisfying the Poisson equation: $\nabla^2\widetilde{\v F}(\v x,t)={\bff}(\v x,t)$. The fields $\v F$ and  $\widetilde{\v F}$ are assumed to vanish at infinity. The implied equality $\Box^2\v F=\nabla^2\widetilde{\v F}$ originates in turn a defined relation between the fields $\v F$ and $\widetilde{\v F}$ by means of which we can always construct one of them by specifying the other one. In particular, if we specify $\widetilde{\v F}$ and solve $\Box^2\v F=\nabla^2\widetilde{\v F}$ then we obtain the solution $\v F=g(\widetilde{\v F})$ which expresses $\v F$ in terms of $\widetilde{\v F}$. Regardless if $\widetilde{\v F}$ is a physical field or not, it can be used to generate  $\v F$. 
It is pertinent to emphasize that the result that $\v F$ can be expressed in terms of 
$\widetilde{\v F}$ is similar to the well-known result that a retarded field can be expressed in terms of its associated (unphysical) potential. The above argument naturally applies to electromagnetism because the retarded electric and magnetic fields satisfy wave equations with confined sources and therefore the {\it formal} existence of instantaneous electric and magnetic fields is actually implied.

In this paper we show how instantaneous electromagnetic fields can be introduced as unphysical objects into classical electrodynamics. Specifically, we obtain the retarded fields of Maxwell's theory in terms of the instantaneous fields of a Galilei-invariant electromagnetic theory [11,12] and we find the specific vector function that transforms the instantaneous fields into the retarded ones.

Let us specify what we understand for instantaneous electric and magnetic fields from a formal point of view. 
Such fields are governed by the field equations [11,12]:
\begin{subequations}
\begin{align}
\nabla\cdot\widetilde{\v E}&= 4\pi\rho,\\ 
\nabla\cdot\widetilde{\v B}&= 0, \\
\nabla\times\widetilde{\v E}&= 0,\\ 
\nabla\times\widetilde{\v B}-\frac{1}{c}\frac{\partial\widetilde{\v E}}{\partial t}&=\frac{4\pi}{c}\v J,
\end{align}
\end{subequations}
where $\widetilde{\v E}$ and $\widetilde{\v B}$ are the instantaneous electric and magnetic fields produced by 
the charge and current densities, $\rho$ and $\v J$. Equations (1) are consistent with the continuity equation and are exactly invariant under Galilei transformations [11,12]. Clearly, eqs. (1) are fundamentally different from Maxwell's equations 
because of the absence of Faraday's induction term [$+(1/c)\,\partial\widetilde{\v B}/\partial t$] in the left-hand side of Eq.~(1c). Of course, this does not mean that $\partial\widetilde{\v B}/\partial t=0$. Time-dependence of the fields $\widetilde{\v E}$ and $\widetilde{\v B}$ is arbitrary because of the assumed arbitrariness of 
$\rho$ and $\v J$. Accordingly, eqs. (1) are not considered here as a limit case of Maxwell's equations [11]. 

Because the fields $\widetilde{\v E}$ and $\widetilde{\v B}$ clearly violate the principle of causality we consider them only as mathematical objects lacking of physical meaning. Therefore, eqs. (1) determine the unobservable fields 
$\widetilde{\v E}$ and $\widetilde{\v B}$ in terms of the observable sources $\rho$ and $\v J$.
The expected relation between instantaneous and retarded fields relies on a formal result according to which 
if we specify the confined sources $\rho$ and $\v J$ then we can always construct simultaneously (i) retarded 
electric and magnetic fields satisfying Maxwell's equations and (ii) instantaneous electric and magnetic 
fields satisfying eqs. (1). 

From  eqs. (1) we derive the Poisson equations:
\begin{subequations}
\begin{align}
\nabla^2\widetilde{\v E}&= 4\pi\nabla\rho,\\
\nabla^2\widetilde{\v B}&=-\frac{4\pi}{c}\nabla\times \v J,
\end{align}
\end{subequations}
with the instantaneous solutions  
\begin{subequations}
\begin{align}
\widetilde {\v E}(\v x,t)&=-\nabla\int \frac{\rho(\v x',t)}{R}\, d^3x',\\
\widetilde {\v B}(\v x,t)&=\nabla\times\int \frac{\v J(\v x',t)}{Rc}\, d^3x',
\end{align}
\end{subequations}
where the integrals are taken over all space and $R=|\v x-\v x'|$. Using the scalar and vector  potentials $\widetilde{\Phi}_L$ and $\widetilde{\v A}_L$ in the Lorenz gauge, $\nabla\cdot\widetilde{\v A}_L+(1/c)\partial \widetilde{\Phi}_L/\partial t=0$, eqs. (3) read
\begin{subequations}
\begin{align}
\widetilde{\v E}&=-\nabla\widetilde{\Phi}_L,\\
\widetilde{\v B}&= \nabla\times \widetilde{\v A}_L,
\end{align}
\end{subequations}
where 
\begin{subequations}
\begin{align}
\widetilde{\Phi}_L(\v x,t)&=\int \frac{\rho(\v x',t)}{R} d^3x',\\
\widetilde {\v A}_L(\v x,t)&=\int\frac{\v J(\v x',t)}{Rc}d^3x'.
\end{align}
\end{subequations}
These potentials satisfy the Poisson equations:
\begin{subequations}
\begin{align}
\nabla^2\widetilde{\Phi}_L&=-4\pi\rho, \\
\nabla^2 \widetilde{\v A}_L&=-\frac{4\pi}{c}{\v J}. 
\end{align}
\end{subequations}

Consider now the Maxwell equations 
\begin{subequations}
\begin{align}
\nabla\cdot\v E&= 4\pi\rho,\\
\nabla\cdot\v B&= 0, \\ 
\nabla\times\v E+\frac{1}{c}\frac{\partial\v B}{\partial t}&=0, \\
\nabla\times\v B-\frac{1}{c}\frac{\partial\v E}{\partial t}&= \frac{4\pi}{c}\v J,
\end{align}
\end{subequations}
and their associated wave equations
\begin{subequations}
\begin{align}
\Box^2\v E&=  4\pi\nabla\rho+\frac{4\pi}{c^2}\frac{\partial\v J}{\partial t},\\
\Box^2\v B&= -\frac{4\pi}{c}\nabla\times \v J.
\end{align}
\end{subequations}
The retarded solutions of  eqs. (8) can be written as
\begin{subequations}
\begin{align}
\v E(\v x,t)&= -\nabla\int \frac{\rho(\v x',t-R/c)}{R}d^3 x'
- \frac 1c\frac{\partial}{\partial t} \int \frac{\v J(\v x',t-R/c)}{Rc}d^3 x',\\
\v B(\v x,t)&= \nabla\times \int \frac{\v J(\v x',t-R/c)}{Rc}d^3 x',
\end{align}
\end{subequations}
where the integrals are taken over all space. Using the scalar and vector  potentials ${\Phi}_L$ and ${\v A}_L$ in the Lorenz gauge, $\nabla\cdot{\v A}_L+(1/c)\partial {\Phi}_L/\partial t=0$, eqs. (9) read
\begin{subequations}
\begin{align}
\v E&= -\nabla\Phi_L- \frac{1}{c}\frac{\partial\v A_L}{\partial t},\\
\v B&= \nabla\times\v A_L,
\end{align}
\end{subequations}
where 
\begin{subequations}
\begin{align}
\Phi_L(\v x,t)&= \int \frac{\rho(\v x',t-R/c)}{R}d^3 x',\\
\v A_L(\v x,t)&= \int \frac{\v J(\v x',t-R/c)}{Rc}d^3 x'.
\end{align}
\end{subequations}
These potentials satisfy the wave equations:
\begin{subequations}
\begin{align}
\Box^2\Phi_L&=  -4\pi\rho,\\
\Box^2\v A_L&= -\frac{4\pi}{c}\v J.
\end{align}
\end{subequations}

We consider now the following theorem: If $\bff(\v x,t)$ and $\bfh(\v x,t)$ represent the confined sources of the retarded field $\v F(\v x,t)$, {\it i.e.}, $\Box^2\v F(\v x,t)=\bff(\v x,t)+ \bfh(\v x,t)$  and if 
$\bff(\v x,t)$ is also the source of the instantaneous field $\widetilde {\v F}(\v x,t)$, {\it i.e.}, 
$\nabla^2\widetilde{\v F}(\v x,t)=\bff(\v x,t)$ then $\v F$ can be obtained from $\widetilde{\v F}$ and $\bfh$
by means of the following relation:
\begin{equation}
\v F= \widetilde{\v F}-\frac{1}{4\pi c^2}\frac{\partial^2}{\partial t^2}\int\int G_R\,\widetilde{\v F} \;d^3x'dt'
-\frac{1}{4\pi}\int\int G_R\,\bfh \;d^3x'dt',
\end{equation}
where the space integrations are over all space and the time integrations is from $-\infty$ to $+ \infty$; 
the fields $\v F$ and $\widetilde{\v F}$ are assumed to vanish at infinity; and the function 
$G_R= \delta(t'-t+R/c)/R$ is the retarded Green function
satisfying $\Box^2 G_R(\v x,t;\v x',t') = -4\pi\delta(\v x-\v x')\delta(t-t')$. 
The proof of the theorem is as follows. Because of $\bff$ is a common source of the fields $\v F$ and 
$\widetilde{\v F}$, it follows that 
\begin{equation}
\Box^2\v F=\nabla^2\widetilde{\v F}+{\bfh}. 
\end{equation}
If $\widetilde{\v F}$ is specified then the solution of eq. (14) can be written 
in the form given by eq. (13). Of course, the D'Alambertian of eq. (13) gives eq. (14).

From eqs. (2a) and (8a) we obtain $\Box^2\v E=\nabla^2\widetilde{\v E} +(4\pi/c^2)\partial \v J/\partial t$ 
and therefore we can apply eq. (13) together with eq. (1c) to obtain, after an integration by parts, 
 the field  $\v E$ in terms of the fields $\widetilde{\v E}$ and $\widetilde{\v B}$: 
\begin{subequations}
\begin{eqnarray}
\v E=\widetilde{\v E}-\frac{1}{4\pi c}\nabla\times\frac{\partial}{\partial t}\int\int G_R\widetilde{\v B}\;d^3x' dt'.
\end{eqnarray}
From eqs. (2b) and (8b) we obtain $\Box^2\v B=\nabla^2\widetilde{\v B}$ and then we can apply eq. (13) 
to find $\v B$ in terms of $\widetilde{ \v B}$:
\begin{eqnarray}
\v B=\widetilde{\v B}-\frac{1}{4\pi c^2}\frac{\partial^2}{\partial t^2}\int\int G_R\,\widetilde{\v B} \;d^3x' dt'.
\end{eqnarray}
\end{subequations}
Clearly, eqs. (15) mix fields having different symmetries: $\v E$ and $\v B$ belong to a Lorenz-invariant theory while $\widetilde{\v E}$ and $\widetilde{\v B}$ belong to a Galilei-invariant theory. {\it Given the charge and current densities we can obtain the instantaneous electric and magnetic fields [via eqs. (3)] and hence the retarded electric and magnetic fields
[via eqs. (15)]:}
\begin{equation}
\{\rho, \v J \}\;\rightarrow \; \{\widetilde{\v E},\widetilde{\v B} \}\;\rightarrow \;\{ \v E,\v B\}.
\end{equation}
Apparently, the instantaneous fields play a similar paper to that of potentials:
$\{\rho, \v J \}\,\rightarrow \,  \{\Phi, \v A\}\,\rightarrow \,\{ \v E,\v B\}$. 
However, the calculation of potentials is more complicated than that of instantaneous fields because 
of the latter do not involve retardation. Moreover, the potentials are gauge-dependent while 
the instantaneous fields are gauge-invariant. 

We can also relate the Lorentz-gauge instantaneous potentials $\widetilde{\Phi}_L$ and $\widetilde {\v A}_L$ to the Lorentz-gauge retarded potentials $\Phi_L$ and $\v A_L$. From eqs. (6) and (12) we get $\Box^2\Phi_L=\nabla^2\widetilde{\Phi}_L$ and $\Box^2\v A_L=\nabla^2\widetilde{\v A}_L$ and therefore we can apply eq. (13) 
to obtain the potentials $\Phi_L$ and $\v A_L$ in terms of the potentials $\widetilde{\Phi}_L$ and $\widetilde {\v A}_L$: 
\begin{subequations}
\begin{align}
\Phi_L&= \widetilde{\Phi}_L-\frac{1}{4\pi c^2}\frac{\partial^2}{\partial t^2}\int\int G_R\,\widetilde{\Phi}_L \;d^3x' dt',\\
\v A_L&= \widetilde{\v A}_L-\frac{1}{4\pi c^2}\frac{\partial^2}{\partial t^2}\int\int G_R\,\widetilde{\v A}_L\;d^3x' dt'.
\end{align}
\end{subequations}
When we take into account eqs. (5), the second terms of eqs. (17) clearly involve two three-dimensional spatial integrals which seems difficult of handling. 
However, these terms can conveniently be transformed. 
After an integration by parts, eqs. (17) can be written as 
\begin{subequations}
\begin{align}
\Phi_L&= \widetilde{\Phi}_L-\frac{1}{c}\frac{\partial{\chi}_L}{\partial t},\\
\v A_L&= \widetilde{\v A}_L-\frac{1}{c^2}\frac{\partial{\bfchi}_L}{\partial t},
\end{align}
\end{subequations}
where 
\begin{subequations}
\begin{align}
\bfchi_L&= \frac {1}{4\pi }
\int\int G_{R}\frac{\partial \widetilde{\v A}_L}{\partial t'}\,d^3x'dt'\\
\chi_L&=  \frac {1}{4\pi c }
\int\int G_R\frac{\partial \widetilde{\Phi}_L}{\partial t'}\,d^3x'dt'.
\end{align}
\end{subequations}
According to eqs. (18), the time derivative of the functions $\chi_L$ and $\bfchi_L$ transform the instantaneous potentials into the retarded potentials. Moreover, the definition of the functions $\chi_L$ and $\bfchi_L$ was made with the premeditated idea that they satisfy the wave equations:
\begin{subequations}
\begin{align}
\Box^2{\chi}_L&=-\frac 1c\frac{\partial \widetilde{\Phi}_L}{\partial t}, \\
\Box^2{\bfchi}_L&= -\frac{\partial \widetilde {\v A}_L}{\partial t}, 
\end{align}
\end{subequations}
together with the continuity-like equation:
\begin{equation}
\nabla\cdot\bfchi_L+\frac{\partial \chi_L}{\partial t}=0.
\end{equation}
If eqs. (5b) and (11b) are used into eq. (18b) then we obtain 
\begin{equation}
-\frac{1}{c^2}\frac{\partial\bfchi_L}{\partial t}= 
\int d^3 x'\frac {1}{Rc}[\v J(\v x',t'=t-R/c)
-\v J(\v x',t)].
\end{equation}
We integrate both sides with respect to {\it ct} to obtain
\begin{equation}
\bfchi_L(\v x, t)= -\int d^3 x'\frac cR \bigg[\int_{t_{\bf 0}}^{t-R/c}dt'\;\v J(\v x',t')-\int_{t_{\bf 0}}^{t}dt'\;\v J(\v x',t')\bigg] +\bfchi_{0}.
\end{equation}
This equation can compactly be written as 
\begin{equation}
\bfchi_L(\v x, t)=-\int d^3 x'\frac cR \int_{t}^{t-R/c}dt'\;\v J(\v x',t') +\bfchi_{0}.
\end{equation}
We change variables by writing $t'=t-\tau$ to obtain
\begin{subequations}
\begin{eqnarray}
\bfchi_L(\v x, t)= \int d^3 x'\frac cR \int_{0}^{R/c}d\tau\;\v J(\v x',t-\tau) +\bfchi_{0}.
\end{eqnarray}
The integration term $\bfchi_{0}$ is {\it a priori} a function of $\v x$ but not of $t$. Actually, $\bfchi_{0}$ is shown to be a constant if we demand finiteness at infinity [9].

Following the same procedure leading to eq. (25a), we use eqs. (5a) and (11a) 
into eq. (18a) to get the analogous expression for $\chi_L$:
\begin{eqnarray}
\chi_L(\v x, t)= \int d^3 x'\frac cR \int_{0}^{R/c}d\tau\;\rho(\v x',t-\tau) +\chi_{0}.
\end{eqnarray}
\end{subequations}
where $\chi_{0}$ is a constant. Therefore, eqs. (19) involving effectively two three-dimensional spatial integrals have been transformed into eqs. (25) which involve one three-dimensional spatial integral and one time integral 
replacing the spatial nonlocality of the respective source with a temporal nonlocality. Equations (25) are 
easier of manipulating than eqs. (19).

The negative of eq. (25b) is essentially the gauge function $\chi_{\cal C}$ that transforms the Lorenz potentials ${\Phi}_L$ and ${\v A}_L$
into the Coulomb potentials ${\Phi}_{\cal C}$ and ${\v A}_{\cal C}$ [9] and therefore $\chi_L$ is essentially the gauge function that transforms the Coulomb potentials into the Lorenz potentials: 
\begin{subequations}
\begin{align}
{\Phi}_L &=  {\Phi}_{\cal C} -\frac{1}{c}\frac {\partial \chi_L}{\partial t},\\
{\v A}_L&=  {\v A}_{\cal C} +\nabla \chi_L.
\end{align}
\end{subequations}
This means that eqs. (18) and (26) are necessarily related. In fact, from eq. (5a) we conclude that $\widetilde\Phi_L=\Phi_{\cal C}$ and consequently eqs. (18a) and (26a) are shown to be identical. 

Insertion of eqs. (18) into eqs. (10) and the use of eqs. 
(21) and (20b) lead to
\begin{subequations}
\begin{align}
\v E&= \widetilde{\v E} -\frac{1}{c}\nabla\times(\nabla\times \bfchi_L),\\
\v B& = \widetilde{\v B} -\frac{1}{c^2}\nabla\times\frac{\partial\bfchi_L}{\partial t}.
\end{align}
\end{subequations}
These relations are the main result of this paper. 
They agree with the implications $\{\rho, \v J \}\to \{\widetilde{\v E}, \widetilde{\v B}, \bfchi_L\}\to \{ \v E,\v B\}$. This means that $\bfchi_L$ is the function whose spatial and temporal derivatives transform the instantaneous fields into the retarded ones. Of course, we can directly verify that eqs. (27) satisfy Maxwell's equations.
 
Equations (18) and (27) illustrate the idea that {\it potentials and fields of the Galilei-invariant electromagnetic theory can be introduced as formal (unphysical) objects into 
classical electrodynamics.} Interestingly, when 
the functions $\bfchi_L$ and $\chi_L$ are separately considered, the former transforms fields of different theories [eqs. (27)] while the latter transforms potentials (in different gauges) of a same theory [eqs. (26)].

On the other hand, using eqs. (18b), (20b) and (21) we can write eq. (27a) as 
\begin{equation}
\v E=\widetilde{\v E}+\frac 1c {\nabla}\frac{\partial \chi_L}{\partial t}
-\frac{1}{c}\frac{\partial{\v A}_L}{\partial t},
\end{equation}
which combines with eq. (26b) to give
\begin{equation}
\v E= \widetilde{\v E}-\frac 1c \frac{\partial \v A_{\cal C }}{\partial t}.
\end{equation}
This relation has recently been verified by Jackson [9] using a novel form of the Coulomb-gauge potential 
$\v A_{\cal C }$.

In conclusion, in this paper we have obtained the retarded fields of Maxwell's theory in terms of the  
instantaneous fields of a Galilei-invariant electromagnetic theory. We have derived the specific function 
$\bfchi_L$ whose spatial and temporal derivatives transform the instantaneous fields 
into the retarded ones.  We have concluded that the instantaneous fields can formally be introduced as unphysical objects into electrodynamics which can alternatively be used to obtain the physical retarded fields.
\vskip 15pt
\centerline{***}
\vskip 15pt
I am most grateful to Prof. R. F. O'Connell for instructive discussions. I am also 	 
grateful to the Fulbright Program for the Scholarship granted to work as visiting professor in 	 
the Department of Physics and Astronomy of the Louisiana State University. I am grateful 	 
to this institution for its hospitality.


\begin{thebibliography}{99}


\bibitem{1}  
Villeco R. A., Phys. Rev. E {\bf 48} (1993) 4008.

\bibitem{2}
Chubykalo A. E., and Smirnov-Rueda R., Phys. Rev. E {\bf 53} (1996) 5373; 
Ivezic T. and Skovrlj L., Phys. Rev. E {\bf 57} (1998) 3680; Chubykalo A. E. and Smirnov-Rueda R., 
Phys. Rev. E {\bf 57} (1998) 3683.

\bibitem{3}
Chubykalo A. E. and Vlaev S. J., Int. J. Mod. Phys. A {\bf 14} (1999) 3789.

\bibitem{4}
Skovrlj L. and Ivezic T, Int. J. Mod. Phys. A {\bf 17} (2002) 2513.

\bibitem{5}
Jackson J. D., Int. J. Mod. Phys. A {\bf 17} (2002) 3975.

\bibitem{6}
Brill O. L. and Goodman B., Am. J. Phys. {\bf 35} (1967) 832. 

\bibitem{7}
Gardiner C. W. and Drummond P. D., Phys. Rev A {\bf 38} (1993) 4397.

\bibitem{8}
Rohrlich F., Am. J. Phys. {\bf 70} (2002) 411.

\bibitem{9}
Jackson J. D., Am. J. Phys. {\bf 70} (2002) 917.

\bibitem{10}
Heras J. A., Am. J. Phys. {\bf71} (2003) 729.

\bibitem{11}
Le Bellac M. and Levy-Leblond J. M., Nuovo Cimento B {\bf 14} (1973) 217.

\bibitem{12}
Jammer M. and Stachel J., Am. J. Phys. {\bf 48} (1980) 5.



\end{thebibliography}
\end{document}